# Beam–Material Interactions

*N.V. Mokhov[1] and F. Cerutti[2]*
[1]Fermilab, Batavia, IL 60510, USA
[2]CERN, Geneva, Switzerland

**Abstract**
This paper is motivated by the growing importance of better understanding of the phenomena and consequences of high-intensity energetic particle beam interactions with accelerator, generic target, and detector components. It reviews the principal physical processes of fast-particle interactions with matter, effects in materials under irradiation, materials response, related to component lifetime and performance, simulation techniques, and methods of mitigating the impact of radiation on the components and environment in challenging current and future applications.

## 1    Introduction

The next generation of medium- and high-energy accelerators for megawatt proton, electron, and heavy-ion beams moves us into a completely new domain of extreme energy deposition density up to 0.1 MJ/g and power density up to 1 TW/g in beam interactions with matter [1, 2]. The consequences of controlled and uncontrolled impacts of such high-intensity beams on components of accelerators, beamlines, target stations, beam collimators and absorbers, detectors, shielding, and the environment can range from minor to catastrophic. Challenges also arise from the increasing complexity of accelerators and experimental set-ups, as well as from design, engineering, and performance constraints.

All these factors put unprecedented requirements on the accuracy of particle production predictions, the capability and reliability of the codes used in planning new accelerator facilities and experiments, the design of machine, target, and collimation systems, new materials and technologies, detectors, and radiation shielding and the minimization of radiation impact on the environment. Particle transport simulation tools and the physics models and calculations required in developing relevant codes, such as FLUKA [3–5], GEANT4 [6–8], MARS15 [9–12], MCNP6 [13, 14], and PHITS [15, 16], are all driven by application. The most demanding applications are the high-power accelerators (e.g., spallation neutron sources, heavy-ion machines, and neutrino factories), accelerator driven systems, high-energy colliders, and medical facilities [2].

This paper gives a brief overview of the principal issues in the field. It is divided into two main sections. The first section is devoted to specific details of interactions of fast particles with matter. The second section characterizes the behaviour of materials under irradiation and highlights related applications at particle accelerators.

## 2    Interactions of fast particles with matter

Electromagnetic interactions, decays of unstable particles, and strong inelastic and elastic nuclear interactions all affect the passage of high-energy particles through matter. The physics of these processes is described in detail in numerous books, handbooks, and reviews (see, for example, Refs. [2, 17–19]).

At high energies, the characteristic feature of the phenomenon is the creation of hadronic cascades and electromagnetic showers in matter due to multiparticle production in electromagnetic and strong nuclear interactions. Because of consecutive multiplication, the interaction avalanche rapidly accrues, passes the maximum and then dies as a result of energy dissipation between the cascade particles and due to ionization energy loss. Energetic particles are concentrated around the projectile axis forming the shower core. Neutral particles (mainly neutrons) and photons dominate with a cascade development when energy drops below a few hundred megaelectronvolts.

The length scale in hadronic cascades is a nuclear interaction length, $\lambda_I$, (16.8 cm in iron), while in electromagnetic showers it is a radiation length, $X_0$, (1.76 cm in iron); see Refs. [17, 18] for definitions and values of these quantities in other materials. The hadronic cascade longitudinal dimension is $(5 \div 10)\lambda_I$, while in electromagnetic showers it is $(10 \div 30)X_0$. It increases logarithmically with primary energy in both cases. Transversely, the effective radius (95% of energy deposited) for a hadronic cascade is about $\lambda_I$, while for electromagnetic showers it is about $2R_M$, where $R_M$ is the Molière radius equal to $0.0265X_0(Z + 1.2)$. Low-energy neutrons coupled to photons propagate for much larger distances in matter around the cascade core, both longitudinally and transversely, until they thermalize down to an energy of the order of a fraction of an electronvolt and possibly undergo radiative capture, still implying the emission of photons of several megaelectronvolts. Muons—created predominantly in pion and kaon decays during cascade development—can travel hundreds and thousands of metres in matter along the cascade axis. Neutrinos—usual muon partners in such decays—propagate even farther, hundreds and thousands of kilometres, until they exit the Earth's surface.

## 2.1 Nuclear reactions: particle and residue production

Hadron production is ruled by non-elastic nuclear reactions. For a sound description of hadron–nucleus (h–A) and nucleus–nucleus (A–A) interactions, one has to rely on a comprehensive understanding of hadron–nucleon (h–N) interactions over a wide energy range as a basic ingredient. Figure 1 shows the total and elastic N–N cross-sections. Below 1 GeV/$c$, the two cross-sections (total and elastic) tend to coincide both for p–p (n–n) and p–n, rapidly increasing with decreasing energy and with about a factor of three difference between p–p and p–n at the low-energy end, as expected on the basis of symmetry and isospin considerations. At high energies, the isospin dependence disappears and the reaction cross-section, given by the difference between total and elastic cross-sections, becomes predominant.

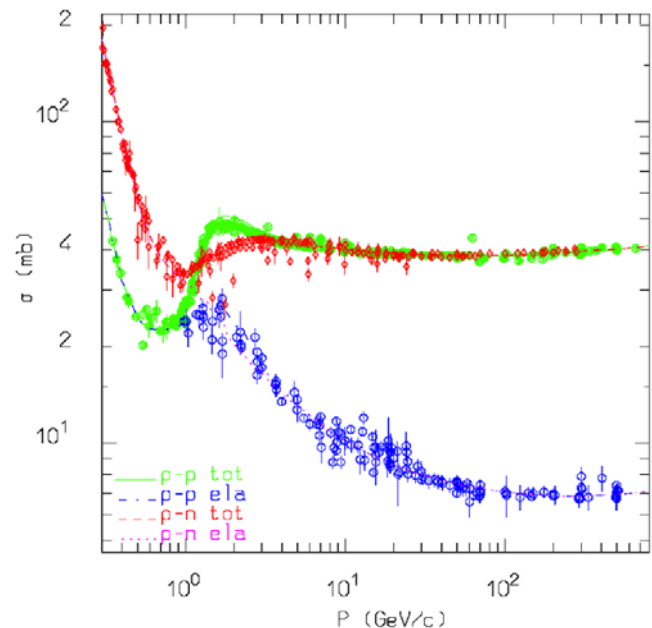

**Fig. 1:** Total (tot) and elastic (ela) proton–proton and proton–neutron cross-sections as a function of proton momentum. Points are experimental data, dashed lines are adopted parameterizations [19].

The non-elastic channel with the lowest threshold, i.e., single pion production, in N–N interactions (N$_1$ + N$_2$ → N$_1'$ + N$_2'$ + π) opens at a projectile kinetic energy of 290 MeV and becomes important above 700 MeV. In pion–nucleon interactions (π + N → π' + π'' + N'), the threshold is as low as 170 MeV. Both reactions are normally described in the framework of the isobar model, assuming that they proceed through an intermediate state containing at least one resonance. There are two main classes of reaction, those in which the intermediate state coincides with a single resonance (possible in π–N reactions) and those in which it initially contains two particles. The former exhibits a bump in the cross-section, corresponding to the mass of the formed resonance. Resonance masses, widths, cross-sections, and branching ratios are extracted from data and conservation laws whenever possible, making explicit the use of spin and isospin relations. They can be also inferred from inclusive cross-sections when needed. For a discussion of resonance production, see, for example, Refs. [20–22].

As soon as the projectile energy exceeds a few gigaelectronvolts, the description in terms of resonance production and decay becomes increasingly difficult. The number of resonances that should be considered grows exponentially and their properties are often poorly known. Furthermore, the assumption of one or two resonance creations is unable to reproduce an experimental feature of high-energy strong interactions, i.e., the large yield of secondary particles that belong neither to the projectile nor to the target fragmentation region but rather to the central region, at small Feynman $x$ values. Different models, based directly on quark degrees of freedom, must be introduced.

Models based on interacting strings have emerged as a powerful tool in understanding quantum chromodynamics at the soft hadronic scale (low transverse momentum), that is in the non-perturbative regime. The dual parton model [23] is one of these models and is built by introducing partonic ideas into a topological expansion, which explicitly incorporates the constraints of duality and unitarity, typical of Regge theory. In this context, hadrons are considered as open strings with quarks, antiquarks, or diquarks sitting at the ends. For instance, mesons (colourless combinations of a quark and an antiquark) are strings with their valence quark and antiquark at the two opposite ends. At sufficiently high energies, the leading term in the interaction corresponds to a pomeron exchange (a closed string exchange), which has a cylindrical topology. When a unitarity cut is applied to the cylindrical pomeron, two hadronic chains are left as the sources of particle production. As a consequence of colour exchange in the interaction, each colliding hadron splits into two coloured partons, one carrying colour charge $c$ and the other $\bar{c}$. The parton with colour charge $c$ (or $\bar{c}$) of one hadron combines with the parton of the complementary colour of the other hadron, to form two colour-neutral chains. These chains appear as two back-to-back jets in their own centre-of-mass systems. The exact method of building up these chains depends on the nature of the projectile–target combination; examples are shown in Fig. 2.

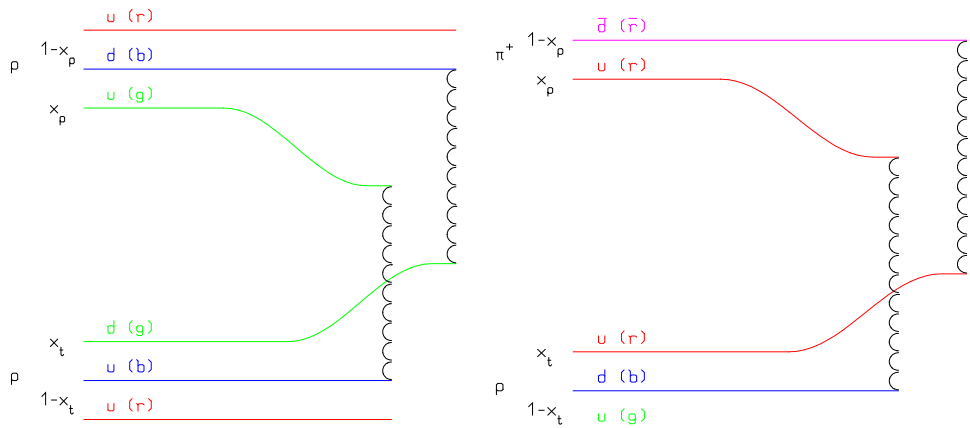

**Fig. 2:** Dual parton model leading two-chain diagram for (left) p–p and (right) π$^+$–p scattering. The respective colour and quark combination shown in the figure is just one of the allowed possibilities. Momentum fractions are also indicated.

The chains produced in an interaction are then hadronized. The dual parton model gives no prescriptions for this stage of the reaction. All the available chain hadronization models, however, rely on the same basic assumptions, the most important one being chain universality; that is, chain hadronization does not depend on the particular process that originated the chain, and until the chain energy is much larger than the mass of the hadrons to be produced, the fragmentation functions (which describe the momentum fraction carried by each hadron) are the same. As a consequence, fragmentation functions can, in principle, be derived from hard processes and $e^+$–$e^-$ data, with (few) parameters valid for all reactions and energies. In fact, mass and threshold effects are non-negligible at typical chain energies and require a suitable treatment. Examples of h–N particle production can be found, for instance, in Ref. [19].

When moving to nucleus interactions (h–A and A–A), the increased complexity of the problem is usually schematized into a sequence of three stages, discussed next.

### 2.1.1 Cascade

In the Glauber formalism [24, 25], the inelastic interaction of a hadron with a nucleus is described through multiple contemporary interactions with $\nu$ target nucleons. The Glauber–Gribov model [26–28] represents the diagram interpretation of the Glauber cascade. The $\nu$ interactions of the hadron projectile originate $2\nu$ chains; two of them are formed by the projectile valence quarks and the valence quarks of one target nucleon (valence–valence chains), while the remaining $2(\nu - 1)$ chains involve projectile sea quarks and valence quarks of the other struck nucleons (sea–valence chains). The chain-building process is illustrated in Fig. 3 for a proton–nucleus interaction; analogous diagrams apply for other hadron projectiles.

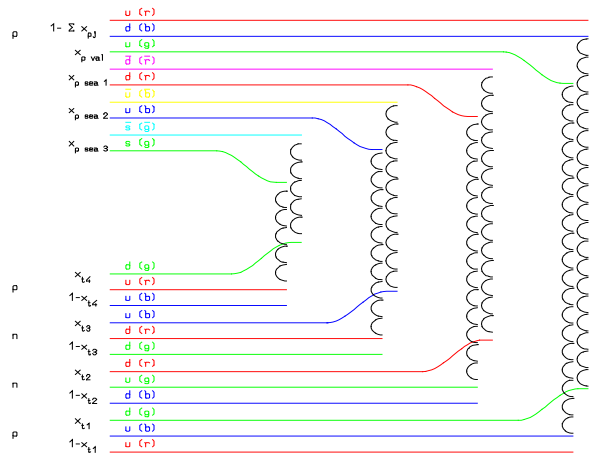

**Fig. 3:** Dual parton model leading two-chain diagram for p–A Glauber scattering with four collisions. The colour and quark combination shown in the figure is just one of the allowed possibilities. Momentum fractions are also indicated.

This sharing of the projectile energy among many chains naturally softens the energy distributions of the reaction products and boosts the multiplicity with respect to h–N interactions. In this way, the model accounts for the major A-dependent features without any degrees of freedom, except in the treatment of mass effects at low energies.

The Fermi motion of the target nucleons must be included to obtain the correct kinematics, in particular, the smearing of transverse momentum ($p_T$) distributions. All nuclear effects on the generated hadrons ('secondaries') are accounted for by the subsequent intranuclear cascade. The formation zone concept is essential to explain the observed reduction of the re-interaction probability with respect to the naive free cross-section assumption. It can be understood as a 'materialization' time. At high energies, the 'fast' particles produced in the Glauber cascade have a high probability of materializing outside the nucleus without triggering a secondary cascade. Further cascading only involves the slow fragments produced in the target fragmentation region of each primary interaction, and therefore the re-interaction probability tends quickly to saturate with energy as the Glauber cascade reaches its asymptotic regime. Only a small fraction of the projectile energy is thus left available for the intranuclear cascade and the following stages. Examples of pion production at different energies are shown in Fig. 4.

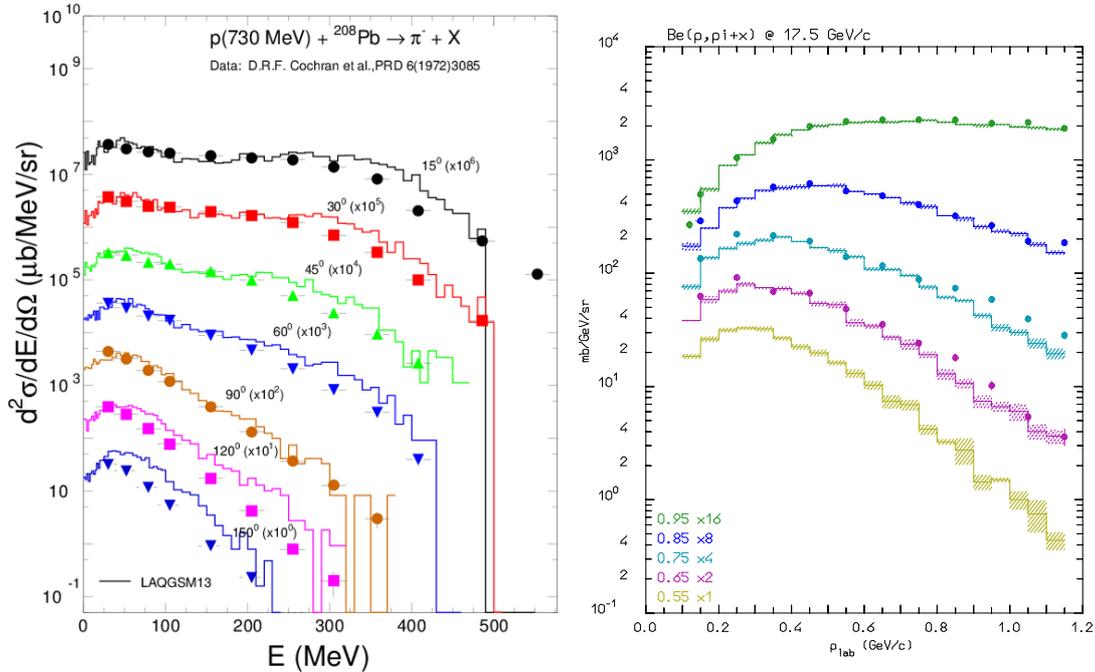

**Fig. 4:** Left: Double differential spectra of negative pions generated by 730 MeV protons on $^{208}$Pb. Symbols are experimental data [29] and histograms are MARS15 results, both scaled according to the angle as indicated. Right: Double differential spectra of positive pions generated by 17.5 GeV/c protons on $^9$Be. Symbols are experimental data [30] and histograms are FLUKA results, both scaled according to the angle as indicated (cosine values are given).

The Glauber cascade and the formation zone act together in reaching a regime where the 'slow' part of the interaction is almost independent of the projectile energy. Owing to the very slow variation of the h–N cross-section from a few gigaelectronvolts to a few teraelectronvolts, the Glauber cascade is almost energy-independent and the rise in the multiplicity of 'fast' particles is related only to the increased multiplicity of the elementary h–N interactions. At the end of the cascading process, the residual excitation energy is directly related to the number of primary and secondary collisions that have taken place. Each collision does indeed leave a 'hole' in the Fermi sea, which carries an excitation energy related to its depth in the Fermi sea.

### 2.1.2 *Pre-equilibrium*

At energies lower than the π production threshold, a variety of pre-equilibrium models have been developed [31], following two leading approaches: the quantum-mechanical multistep model and the exciton model. The former has a very good theoretical background but is quite complex, while the latter relies on statistical assumptions, and is simple and fast. Exciton-based models are often used in Monte Carlo codes to link the intranuclear cascade stage of the reaction to the equilibrium one.

Typically, the intranuclear cascade stage continues until all nucleons in the nucleus are below a smooth threshold of a few tens of megaelectronvolts and all particles except nucleons (e.g., pions) have been emitted or absorbed. The input configuration for the pre-equilibrium stage is characterized by the total number of remaining protons and neutrons, by the number of particle-like excitons (nucleons excited above the Fermi level) and of hole-like excitons (holes created in the Fermi sea by the intranuclear cascade interactions), and by the excitation energy and momentum of the resulting nucleus. All these quantities can be derived by properly recording what occurred during the intranuclear cascade stage.

The pre-equilibrium stage, while distributing the excitation energy among all degrees of freedom through N–N elastic scattering, accounts for intermediate energy emissions of single nucleons and light particles formed by nucleon coalescence.

### 2.1.3 Final de-excitation

The last stage of the reaction chain assumes that the remaining nucleus (typically more than one in the case of A–A interactions, featuring both a projectile- and a target-like residual system) is a thermally equilibrated system, characterized by its mass and atomic numbers and a given excitation energy. The latter is dissipated through the 'evaporation' of single nucleons, light particles, and intermediate mass fragments, or by fission. The neutron evaporation is favoured over charged particle emission, owing to the Coulomb barrier, especially for medium-heavy nuclei, whose excitation energy is higher, owing to the larger cascading chances and to the larger number of primary collisions in the Glauber cascade at high energies.

Many evaporation or fission models are based on the standard Weisskopf–Ewing formalism [32]. For light residual nuclei, where the excitation energy may overwhelm the total binding energy, statistical fragmentation models (Fermi break-up) are more appropriate.

The end of the de-excitation process is characterized by the emission of γ-rays, corresponding to the transition between specific levels of the residual nucleus.

Although the reaction may originate from a particularly high-energy projectile, its evaporation, fission, or break-up stage is a low-energy phenomenon, much slower than the previous stages and sensitive to nuclear physics ingredients. In fact, it determines what is left after the interaction, yielding the distribution of residues, which, in the case of an energetic nuclear reaction on a high-Z material, fills the whole mass (and charge) range of the nuclide chart, as demonstrated in Fig. 5.

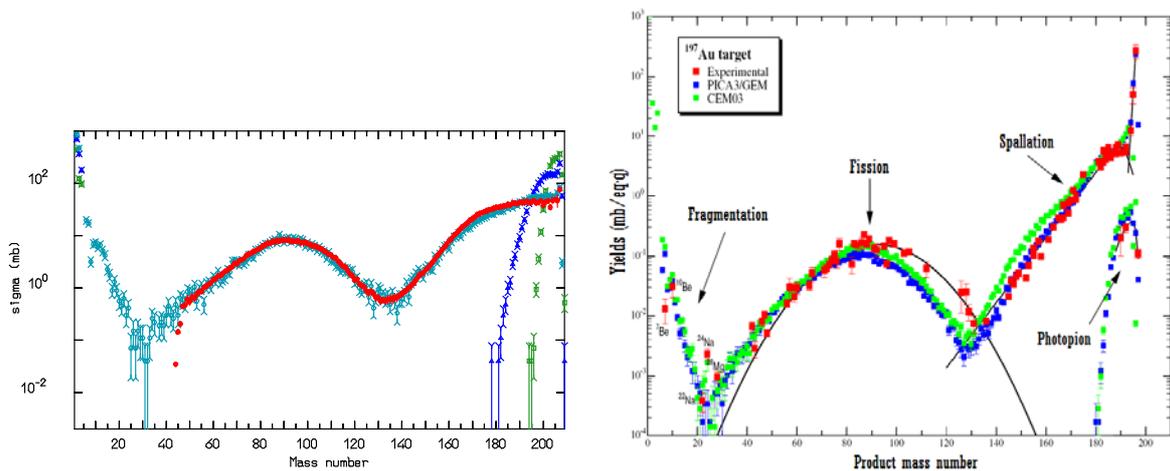

**Fig. 5:** Left: Mass distribution of the nuclei generated by 1 GeV/n $^{208}$Pb ions on hydrogen. Data [33] (red) are compared with FLUKA results (light blue). The distributions obtained after the cascade (green) and pre-equilibrium (dark blue) stages are also shown. Right: Mass distribution of the nuclei generated by bremsstrahlung photons (up to 1 GeV) on $^{197}$Au. Data (see Ref. [34] and references therein) in red are compared with the results of the CEM03 model used in MARS15 (green). The distribution obtained using a different model (blue) is also shown. Reproduced with permission from S. Mashnik.

It is worth mentioning that photonuclear (see Fig. 5 right) and electronuclear interactions can also be coherently described in this framework, through an appropriate definition of the initial state. Electronuclear interactions are, in fact, nuclear interactions by virtual photons.

## 2.2 Radionuclides

Among the residual nuclei generated in non-elastic nuclear reactions, one can probably find radionuclides, which are subject to further decay into other nuclear species and so are responsible for continuous delayed radiation, mostly of an electromagnetic nature (electrons, positrons, and photons), owing to β and γ decays. This initiates decay chains, governed by the radioactivity laws, where the time evolution of isotope populations is given by the Bateman equations:

$$\mathrm{d}N_n/\mathrm{d}t = P_n + \sum_i (b_{i,n} \cdot \lambda_i \cdot N_i) - \lambda_n \cdot N_n \ . \tag{1}$$

In Eq. (1), $N_n$ represents the population of a certain isotope, $n$, varying as a function of its direct production rate by nuclear reactions, $P_n$, its decay constant, $\lambda_n$, and the growth rate by parent decay expressed by the sum extended to all parent isotopes, $i$. The latter takes into account the respective branching ratios $b_{i,n}$ for the relevant channel.

A Monte Carlo code simulating the *prompt radiation* propagation in a given geometry, and thereby calculating the radionuclide production rates, $P_n$, can also be provided with the capability of solving, on-line, the system of Bateman equations for an irradiation profile (sequence of time intervals and corresponding beam intensities) and several cooling times (time distances from the irradiation end). This enables it to calculate specific activities as well as residual dose rates owing to the *decay radiation*, whose spatial propagation can be simulated at the same time.

## 2.3 Ionization energy loss

Tracking the transport of charged particles in matter involves accounting for the Coulomb interactions they experience with the electrons and the atomic nuclei of the medium. While the incoming particle energy loss is largely dominated by the first interactions, the other interactions are the main responsible for the particle's trajectory changes. In most cases, multipurpose Monte Carlo codes treat these processes as *continuous*, contrary to *discrete* events, such as nuclear reactions, bremsstrahlung emissions, Compton scattering, or particle decay. This means that the particle proceeds through steps, at the end of each of which its energy and direction is changed to take into account the cumulative effect of all Coulomb interactions (inducing atom ionization and excitation) experienced along the track. Actually, the generation of energetic knock-on electrons (δ rays) can also be treated as a discrete event (above the electron transport energy limit, of the order of 1 keV), paying a significant penalty in computation time, where it is justified by the need not to assume the particle energy loss as translating into a local energy deposition, but to consider the range of the electron carrying part of that energy loss elsewhere. Analogously, as far as the particle trajectory is concerned, single scattering with an atomic nucleus can also be explicitly simulated when needed (implying much shorter steps), instead of utilizing multiple scattering algorithms, providing the trajectory alteration as a cumulative outcome.

The well-known Bethe–Bloch formula [35] gives the *mean* loss rate (stopping power) as a function of the particle speed and charge and of the relevant material properties. Nevertheless, several corrections (such as shell, Barkas, Bloch, Mott, and Lindhard–Sørensen corrections) have to be included, to ensure suitable accuracy. Moreover, with high-$Z$ projectiles it is necessary to evaluate their effective charge, since electron capture becomes important at low energies. In addition, actual energy losses feature significant fluctuations with respect to the mean value, making the latter far from being exhaustive, but requiring the implementation of a proper distribution function.

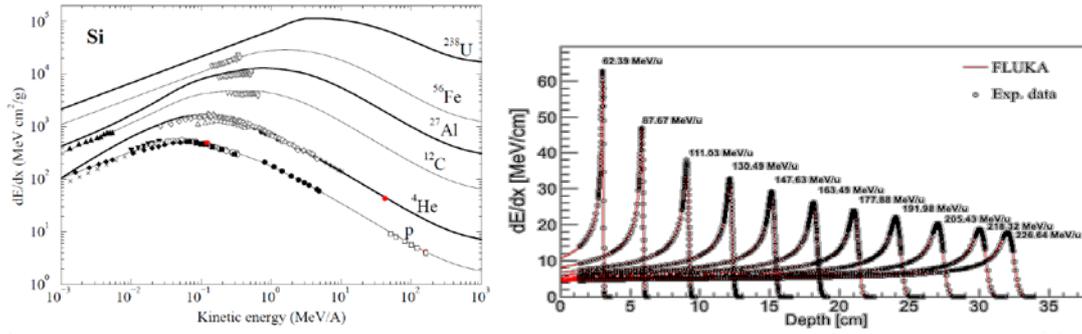

**Fig. 6:** Left: Stopping power of several projectile species in silicon. Symbols are measured values [36–41] and curves are MARS15 predictions. Right: Profiles of energy deposition in water by protons at different energies in the hadron therapy range. Circles are experimental data and curves are FLUKA results [42].

Examples of the accuracy achieved in the description of the ionization process are shown in Fig. 6, where, in addition to the reproduction of measured stopping power values for a notable variety of radiation types, the study of many Bragg peaks of clinical use in proton therapy is reported.

## 2.4 Displacements of atoms

The dominant mechanism of structural damage of inorganic materials is displacement of atoms from their equilibrium position in a crystalline lattice as a result of irradiation, with the formation of interstitial atoms and vacancies in the lattice. The resulting deterioration of material critical properties is characterized—in the most universal way—as a function of the number of displacements per target atom; this number is a strong function of projectile type, energy, and charge, as well as material properties, including temperature.

Three major codes (FLUKA, MARS15, and PHITS) use very similar implementations of the Norgett-Robinson-Torrens (NRT) model [43, 44] to calculate the number of displacements per target atom [2]. A primary knock-on atom created in nuclear collisions can generate a cascade of atomic displacements. This is taken into account via the damage function $v(T)$. The number of displacements per target atom is expressed in terms of the damage cross-section $\sigma_d$:

$$\sigma_d(E) = \int_{T_d}^{T_{max}} \frac{d\sigma(E,T)}{dT} v(T) dT , \qquad (2)$$

where $E$ is the kinetic energy of the projectile, $T$ is the kinetic energy transferred to the recoil atom, $T_d$ is the displacement energy, and $T_{max}$ is the highest recoil energy according to kinematics. In a modified Kinchin–Pease model [43], $v(T)$ is zero at $T < T_d$, unity at $T_d < T < 2.5T_d$, and $k(T)E_d/2T_d$ at $2.5T_d < T$, where $E_d$ is 'damage' energy available to generate atomic displacements by elastic collisions. $T_d$ is an irregular function of atomic number (~40 eV). The displacement efficiency, $k(T)$, introduced as a result of simulation studies on evolution of atomic displacement cascades [45], drops from 1.4 to 0.3 once the primary knock-on atom energy is increased from 0.1 to 100 keV, and exhibits a weak dependence on target material and temperature.

The implementation of this model in MARS15 [46] and FLUKA [47] includes electromagnetic elastic (Coulomb) scattering, the Rutherford cross-section with Mott corrections, and nuclear form-factors (a factor of two effect). Resulting displacement cross-sections due to Coulomb scattering are shown in Fig. 7 (left) for various projectiles on silicon and carbon targets. For elementary particles, the energy dependence of $\sigma_d$ disappears above 2–3 GeV, while it continues to higher energies for heavy ions. For projectiles heavier than a proton, $\sigma_d$ grows with the projectile charge $z$ as $z^2/\beta^2$ at $\gamma\beta > 0.01$, where $\beta$ is the projectile velocity. All products of elastic and inelastic nuclear interactions, as well as Coulomb elastic scattering of transported charged particles (hadrons, electrons, muons, and heavy ions) from 1 keV to 10 TeV, contribute to the number of displacements per target atom in the model. The number of displacements per target atom for neutrons from $10^{-5}$ eV to 20–150 MeV is described in MARS15 using the NJOY99+ENDF-VII database [48, 49] for 393 nuclides [50]. A corresponding output is shown in Fig. 7 (right). FLUKA adopts database information for neutrons up to 20 MeV, while at higher energies, where many reaction channels are open, it describes neutron elastic and inelastic interactions through its models and determines the number of displacements per target atom explicitly by calculating non-ionizing energy losses of the products.

Such results are then corrected using the experimental defect production efficiency $\eta$, where $\eta$ is a ratio of a number of single interstitial atom vacancy pairs (Frenkel pairs) produced in a material to the number of defects calculated using the NRT model. The values of $\eta$ have been measured [51] for many important materials in the reactor energy range.

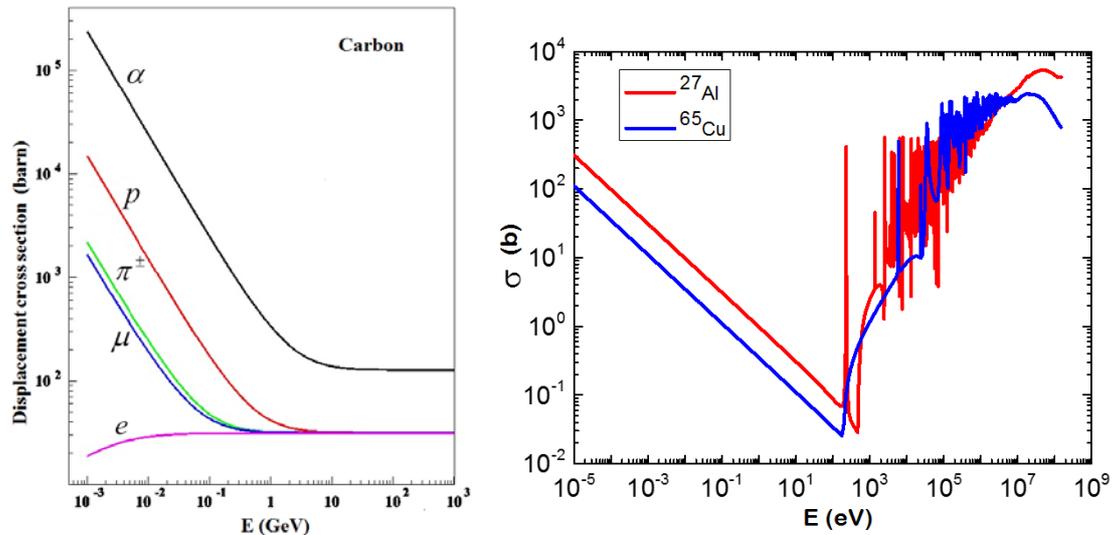

**Fig. 7:** Left: Displacement cross-section in carbon for various charged particles (MARS15). Right: NRT neutron defect production cross-sections on aluminium and copper.

## 3    Materials under irradiation

Depending on the material, the level of energy deposition density, and the time structure, one can face a variety of effects in materials under irradiation. The two categories of materials response are related to the component lifetime and performance [2]:

1. component damage (lifetime):
   - thermal shocks and quasi-instantaneous damage (see also A. Bertarelli's contribution in these proceedings);
   - insulation property deterioration due to dose build-up;
   - radiation damage to inorganic materials due to atomic displacements, as well as helium and hydrogen production;
   - detector component radiation aging and damage.

2. operational (performance):
   - superconducting magnet quenching;
   - single-event effects in electronics;
   - detector performance deterioration;
   - radioactivation, prompt dose and impact on environment.

## 3.1 Thermal shock

Short pulses with energy deposition density ranging from 200 J/g (tungsten) and 600 J/g (copper) to ~1 kJ/g (nickel, Inconel) and ~15 kJ/g cause thermal shocks, resulting in fast ablation and slower structural changes, or melting. Two outstanding examples are the Fermilab antiproton ($\bar{p}$) production target damages by a 120 GeV proton beam in the period from 1993 to 2011, and a 980 GeV proton beam-induced Tevatron collimator damage.

### 3.1.1 Antiproton production target

Significant effects in the evolution of the Fermilab antiproton production target have been observed [52]. This 10 cm diameter target stack is made up of six target discs 0.95 cm thick separated—in early days—by two 0.32 cm thick copper cooling discs, later replaced with copper mini-cylinders to provide better airflow. The entire assembly slowly rotates, distributing the primary beam, with time, over a cylindrical section of the target. In Tevatron Run I at Fermilab, evidence of external target damage, sustained when the rotation mechanism failed for several months with only vertical motion available, was discovered. Figure 8 shows damage at the exit of a nickel target chord. Ejection of nickel pieces has also led to a contamination incident.

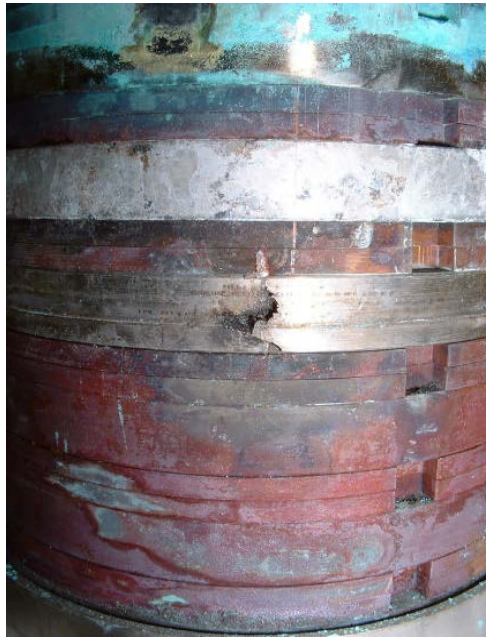

**Fig. 8**: Tevatron Run I antiproton production target damage in 1994. Courtesy of A. Leveling and J. Morgan

When the target was rotated properly, it was not damaged, although the outer titanium sleeve showed signs of swelling. Nickel was used in the first year of Run II. After several months of operation at $4.5 \times 10^{12}$ protons per pulse with root mean square beam spot sizes of $\sigma_x = 0.22$ mm and $\sigma_y = 0.16$ mm, a region of damage about 2.5 mm wide developed on the target; the titanium jacket evaporated in that region and there was a 15% drop in $\bar{p}$ yield.

In September 2002, the targets were replaced with Inconel targets, which have an excellent high-temperature tensile strength, although a relatively low thermal conductivity compared with copper and nickel. The switch to Inconel-600 extended the service life of each target from weeks to months (×6 for the entire stack), with practically no decrease in $\bar{p}$ yield. The 11.43 cm outer diameter Inconel target disc with the copper mini-cylinders, providing best airflow for cooling, is shown in Fig. 9 (left).

In 2007, the target was used in a 'consumable mode' to maximize the $\bar{p}$ yield. Figure 9 (right) shows the Inconel-600 disc remnants after 4 months of operation with a total $2.65 \times 10^{19}$ protons on target and a root mean square beam spot size of $\sigma_x \sim 0.18$ mm, $\sigma_y \sim 0.22$ mm. The observed phenomenon was attributed to chemical oxidation of the damaged—by thermal shock—target surface at the target chord beam exit. Because of tight limits established for vertical target positioning, the copper cooling cylinders between the target discs emerged relatively undamaged, though distorted in some places, as seen in the figure.

Radioactive particles from a damaged target were also a problem amplified by radioactive titanium pieces ejected from the damaged jacket. In Run II, the titanium jacket was replaced by a thin-walled cylinder of graphite, and then by a 6.35 mm thick beryllium jacket, both being nearly transparent to the primary beam and products of its interactions with the target.

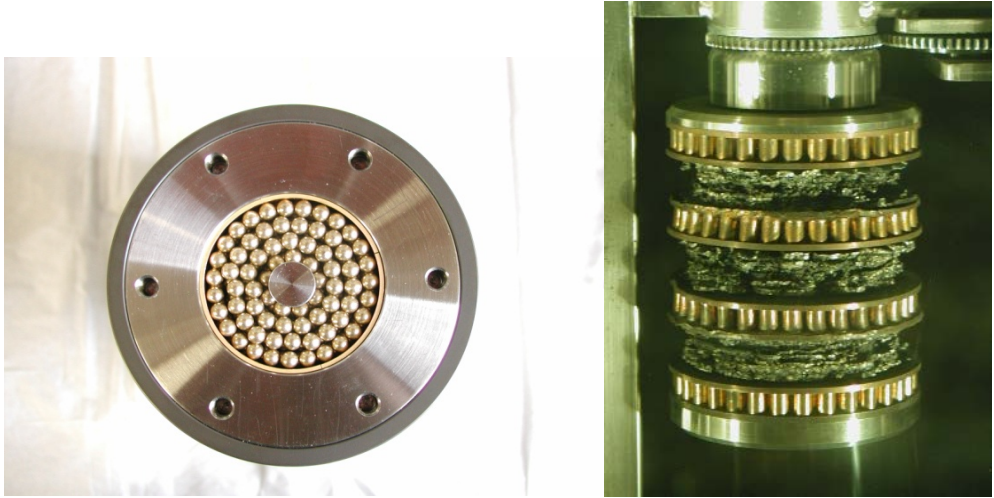

**Fig. 9:** Tevatron Run II antiproton production target. Left: Top view. Right: Target number 7 in 2007 with damaged Inconel discs and distorted copper cooling cylinders in between. Courtesy of A. Leveling and J. Morgan.

### *3.1.2 Tevatron collimator damage*

Another example of the fast material ablation at accelerators is the destruction of the Tevatron primary (Fig. 10, left) and secondary (Fig. 10, right) collimators caused by an accidental loss of the 980 GeV beam in 2003 [53]. The damage was induced by a failure in the Collider Detector Facility Roman pot detector positioning at the end of a 980 × 980 GeV proton–antiproton colliding beam store.

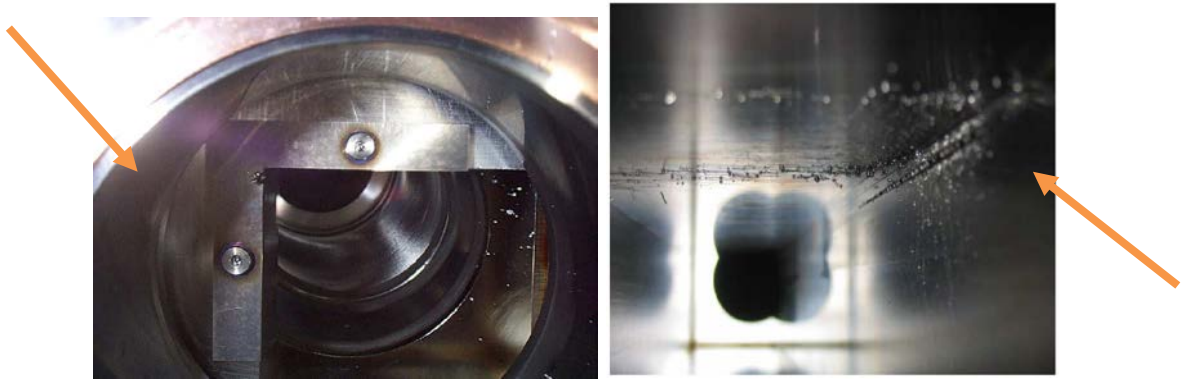

**Fig. 10:** Left: The hole indicated was created in the Tevatron 5 mm thick primary tungsten collimator. Right: The 25 cm long groove indicated was created in the Tevatron secondary stainless steel collimator [2, 53].

The dynamics of this failure over the first 1.6 ms, including excessive halo generation and superconducting magnet quenching, were studied via realistic simulations using the MARS [9–12] and STRUCT [54] codes. It was shown that the misbehaved beam-induced ablation of the tungsten primary collimator resulted in the creation of the hole seen in it, while the simulated parameters of the groove in the stainless steel secondary collimator jaw surface fully agreed with the post-mortem observations [2, 53].

## 3.2 Hydrodynamic regime

Beam pulses with energy deposition density in excess of 15 kJ/g bring materials to the hydrodynamic regime [2]. This was demonstrated in studies [55] for the SSC 20 TeV proton beam (400 MJ, 300 µs spill), first on a graphite beam dump and later for the collider superconducting magnets, steel collimators, and tunnel-surrounding Austin chalk. Since the beam duration was comparable to the characteristic time of expected hydrodynamic motions, the static energy deposition capability of the MARS code has been combined with the 2D and 3D hydrodynamics of the LANL's MESA and SPHINX codes. It was found, in simulations, that a hole was drilled by the beam in the graphite dump at a rate of 7 cm/µs with generated pressures of a few kbar (Fig. 11, left). Later these effects were studied in detail for the Super Proton Synchrotron (SPS) and Large Hadron Collider (LHC) targets, beam dumps, and collimators using coupling of the FLUKA code (energy deposition) with BIG2 [56, 57] and LS-DYNA [58] (hydrodynamics) codes. Figure 11 (right) shows the calculated physical state of the solid tungsten target at the end of the SPS proton pulse (root mean square beam spot size of 0.088 mm) at 7.2 µs. It can be seen that within the inner 2 mm radius, a strongly coupled plasma state exists, which is followed by an expanded hot liquid. The melting front is seen propagating outwards.

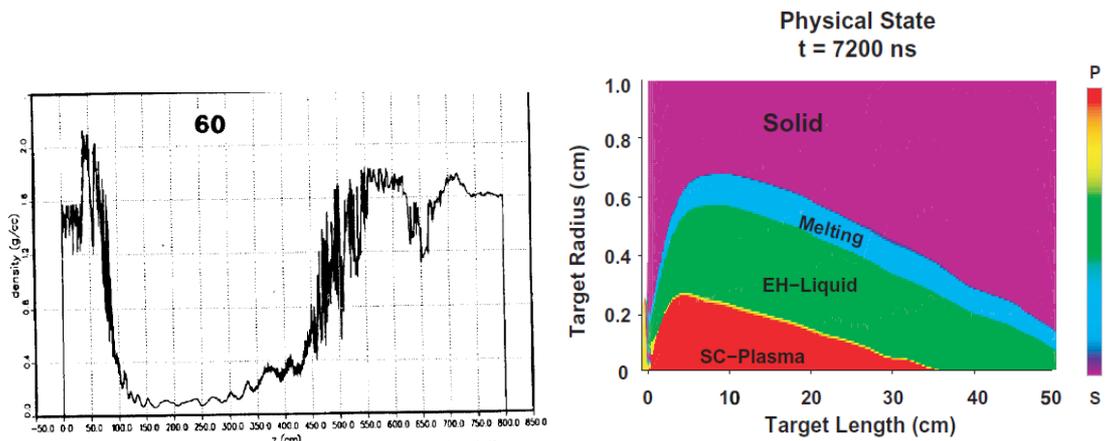

**Fig. 11:** Left: Axial density of graphite beam dump in 60 µs after the 20 TeV beam spill start [55]. Right: Tungsten target physical state after the SPS beam pulse [56]. EH, expanded hot; SC, strongly coupled. Reproduced from [56] with permission from N. Tahir.

## 3.3 Hydrogen and helium gas production

At accelerators, radiation damage to inorganic structural materials—being primarily driven by displacement of atoms in a crystalline lattice (see Section 2.4)—is amplified by increased hydrogen and helium gas production for high-energy beams. In the Spallation Neutron Source (SNS) beam windows, the ratio of He atoms to the number of displacements per target atom is about 500 times that in fission reactors. These gases can lead to grain boundary embrittlement and accelerated swelling. In the simulation codes analysed here, uncertainties on production of hydrogen are about 20%, while for helium uncertainties could be as high as 50%.

## 3.4 Dose in organic materials

Unlike megaelectronvolt-type accelerators, which have insulators made mostly of ceramics or glasses, the majority of insulators in high-energy accelerator equipment are made of organic materials: epoxy resin, G11, polymers, etc. [2]. Apart from electronics and optical devices, the organic materials are the most sensitive to radiation. A large number of radiation tests have been made on these materials and the results are extensively documented [59]. The impact of radiation on organic materials is a three-step process [60]:

1. production of free radicals by radiation;
2. reaction of free radicals: crosslinking, chain scission, formation of unsaturated bonds (C=C, etc.), oxidation, and gas evolution;
3. change of molecular structure: modification and degradation affected by irradiation temperature and atmosphere as well as by presence of additives.

The findings for organic materials under irradiation are [60]:

– degradation is enhanced at high temperatures;
– radiation oxidation in the presence of oxygen accelerates degradation;
– radiation oxidation is promoted in the case of low dose rates;
– additives can improve radiation resistance; for example, 1% by weight of antioxidant in polyethylene can prolong its lifetime 5 to 10 times.

Dose limits on insulators are usually defined for a certain level of change in the material properties critical to the application. For example, 10% degradation in ultimate tensile strength is a typical criterion for epoxy, CE epoxy resins and G11. Similar changes in electrical resistivity are often used as a criterion. For the given insulator and irradiation conditions, radiation damage is proportional to the peak energy deposition density or dose accumulated in the hottest region. Radiation damage thresholds, based on the results of dedicated radiation tests [59], experience, or indirect evidence, are used worldwide as a basis for design and an estimate of component lifetime or operation time prior to replacement. For example, the dose limit used for the LHC superconducting magnet insulators is 25 to 40 MGy [61]. Other projects utilizing superconducting magnet technologies assume a lower limit, of 7–10 MGy.

It is worth noting here that energy deposition—responsible for damage in insulators and, e.g., for cable quench stability in superconducting magnets—is modelled in accelerator applications quite accurately. In the majority of cases, FLUKA and MARS15 results on energy deposition coincide within 10% and agree with data.

### 3.5 Quenching

A magnet quench is a dramatic yet fairly routine event within a particle accelerator. It occurs when one of the superconducting magnets that steer and focus the particle beams warms above a critical temperature, bringing operations to an abrupt halt. A quench often starts when stray particles from the beam enter a magnet's coils, producing an initial burst of heat. Within a fraction of a second, parts of the superconducting wire in the magnet lose their ability to conduct electricity without resistance, generating more heat that quickly spreads throughout the entire magnet. The coolant surrounding the magnet begins to boil.

As with the other types of the beam impact on materials, the beam-induced quench creation and propagation in the superconducting coils depend on the level and profile of energy deposition density, its time structure, operational current and—for pulses longer than ~1 ms—on the cooling efficiency of the superconducting cable. The accelerator class superconducting magnet will quench if the peak power density in the innermost cable exceeds 40–60 mW/cm$^3$ in the Nb$_3$Sn quadrupoles at $I_{op}/I_c = 0.8$ [62–64]. Here, $I_{op}$ is the magnet operational current, and $I_c$ is the magnet critical current. The quench limit in NbTi based coils is 13 mW/cm$^3$, again at $I_{op}/I_c = 0.8$. Studies of beam-induced effects in accelerator superconducting magnets are described in Refs. [61, 64], which consider the high-luminosity upgrade of the LHC inner triplet magnets. A tiny fraction of the 7 TeV proton beams or products of their interactions lost on the superconducting magnets would induce hadronic and electromagnetic showers with energy deposition levels that could easily exceed these quench limits. Optimized using thorough

FLUKA and MARS15 studies, absorbers and high-$Z$ magnet bore inserts (related to the electromagnetic nature of energy deposition at that location) are to be incorporated in the high-luminosity LHC inner triplet region to mitigate this problem safely. Figure 12 from Ref. [61] shows the details of the protection system (left) and the resulting peak power density profile—well below the quench limits—in the innermost superconducting cable by the collision debris at the luminosity of $5 \times 10^{34}$ cm$^{-2}$ s$^{-1}$ (right).

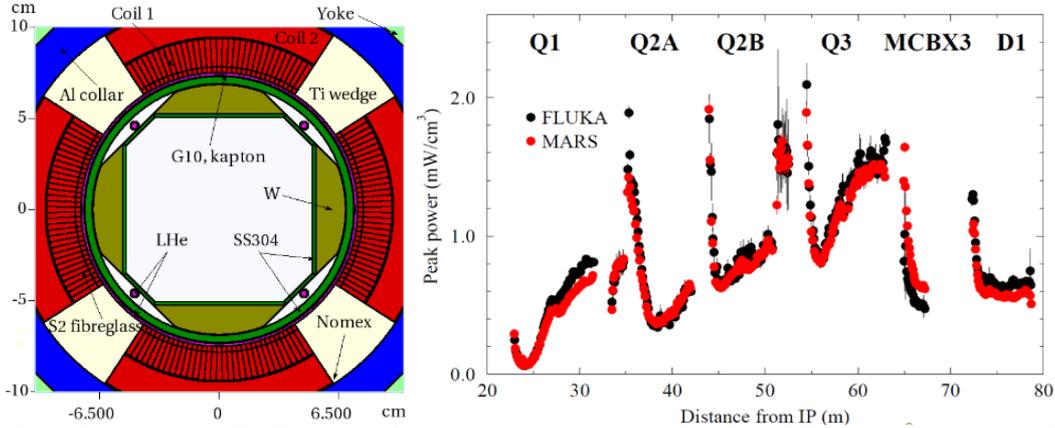

**Fig. 12:** Left: Details of the FLUKA-MARS15 model in the innermost region of the high-luminosity LHC inner triplet first quadrupole, with 16 mm thick tungsten inserts (olive) in the mid-planes. Right: Longitudinal peak power density profile on the innermost superconducting cable of the inner triplet, orbit correctors (MCBX) and separation dipole (D1) coils, as calculated by FLUKA (black) and MARS15 (red) for 14 TeV centre-of-mass collision debris at $5 \times 10^{34}$ cm$^{-2}$ s$^{-1}$ luminosity.

In the case of a large superconducting magnet, which can be several metres long and carry currents of 10,000 A or more, the quench creates a loud roar as the coolant—liquid helium with a temperature close to absolute zero—turns into gas and vents through pressure relief valves, like steam escaping a tea kettle [65]. Such a quench generates as much force as an exploding stick of dynamite. A magnet usually withstands this force and is operational again in a few hours after cooling back down. If repair is required, it takes valuable time to warm up, fix, and then cool down the magnet, i.e., days or weeks in which no particle beams can be circulated, and no science can be done.

During CERN's LHC start-up operations in 2008, with current ramping up and no beam circulating, an electrical fault in a dipole–quadrupole interconnection was responsible for the development of an electrical arc puncturing the helium enclosure [66]. As a consequence, several superconducting magnets quenched and, despite helium relieving to the tunnel, large pressure forces displaced dipoles in a few subsectors. Eventually, the replacement of a number of magnets was necessary. To mitigate potentially destructive quenches, the superconducting magnets that form the LHC are equipped with fast-ramping heaters, which are activated once a quench event is detected by a complex quench protection system. Since the dipole bending magnets are connected in series, each power circuit includes 154 individual magnets; should a quench event occur, the entire combined stored energy of these magnets must be dumped at once. This energy is transferred into dumps that are massive blocks of metal, which heat up to several hundreds of degrees Celsius, through resistive heating, in a matter of seconds.

### 3.6 Radiation to electronics

Electronic components and systems exposed to a mixed radiation field experience three different types of radiation damage: damage from the total ionizing dose, displacement damage, and so-called single-event effects. The latter events range from single event or multiple bit upsets and single-event transients to possible destructive latch-ups, destructive gate ruptures or burn-outs (single-event gate ruptures and burn-outs).

The first two types of damage refer to the steady accumulation of defects causing measurable effects, which can ultimately lead to device failure. They are evaluated through total ionizing dose (in grays) and non-ionizing energy deposition, respectively. The latter is generally quantified by accumulated silicon 1 MeV equivalent neutron fluence, which requires the use of conversion factors to weight the effect of other energies and particle types with respect to one of the 1 MeV neutrons in silicon, as electronic device material. As for stochastic single-event-effect failures, these form an entirely different group, since they result from the ionization by a single particle, which is able to deposit enough energy to perturb the operation of the device. They can only be characterized in terms of their probability of occurrence as a function of accumulated high-energy hadron fluence, not overlooking the dependence on device type as well as on particle nature. Actually, the hadron energy threshold is usually intended as 20 MeV, but unstable hadrons of lower energies must also be counted. Concerning neutrons of lower energies, one has to weight them according to the ratio of their single-event upset cross-section to that of hadrons above 20 MeV, substantially reflecting the ($n$, $x\alpha$) cross-section behaviour in representative microchip materials.

Such failures can lead to serious consequences: for instance, single-event effects were responsible for the shut-down of the CERN Neutrinos to Gran Sasso (CNGS) facility in 2007. The CNGS facility was designed to produce an intense muon neutrino beam directed towards the Gran Sasso National Laboratory (LNGS) in Italy, 732 km from CERN. The physics program extended over 5 yr with up to $4.5 \times 10^{19}$ protons impinging on the primary target per year, extracted from the CERN SPS with two nominal extractions of $2.4 \times 10^{13}$ protons at 400 GeV/$c$ each, in a CNGS cycle of 6 s, corresponding to an average power at the target of 510 kW. The facility was started with gradually increasing intensity in 2007, but had to be shut down after only $8 \times 10^{17}$ protons on target, owing to successive failures in the ventilation system. After detailed analysis, it was found that the microcontrollers that failed were placed in relatively high radiation areas, i.e., near to the ducts connecting the target and service galleries (see Fig. 13). The failure was due to single-event effects induced by high-energy hadrons.

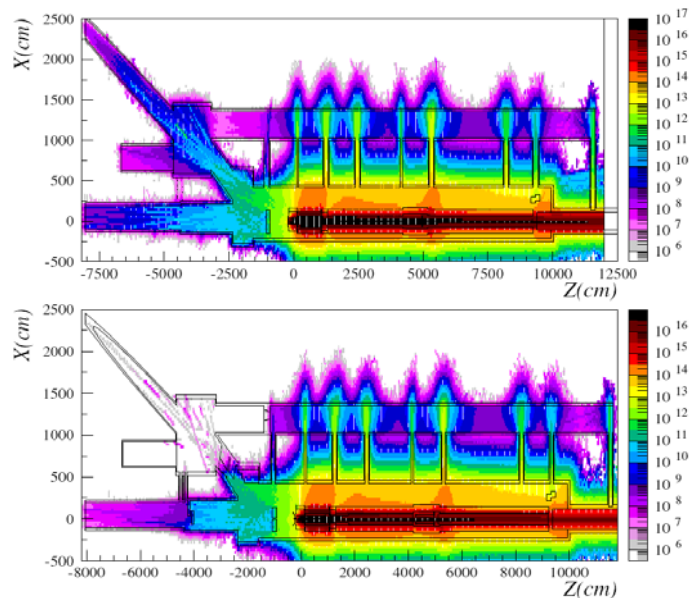

**Fig. 13:** Annual high-energy hadron fluence (in cm$^{-2}$) in the CNGS facility, as predicted by FLUKA calculations, before and after the installation of the shielding aimed to create a protected area (the rectangle centred at about $Z = -20$ m, $X = 12$ m) for the control electronics. The 400 GeV/$c$ proton beam impinges from the left on the carbon target at the reference frame origin. In the mentioned area, the radiation levels, initially in the range of $10^7$ to $10^9$ cm$^{-2}$ yr$^{-1}$, are reduced by the shielding below $10^6$ cm$^{-2}$ yr$^{-1}$.

Enhancement of the electronics protection was mandatory, therefore, a radiation safe area was created with the introduction of fixed and mobile shielding. In total, 53 m$^3$ of concrete was poured *in situ*, the ventilation system had to be completely reconfigured, and all the electronics had to be moved to the new safe area. According to calculations (as in Fig. 13), the new shielding ensures considerable attenuation with respect to the former layout, for all quantities of interest, i.e., total ionizing dose, silicon 1 MeV neutron equivalent (1 MeV n-eq) fluence, and high-energy hadron fluence. In particular, the latter quantity is reduced to only at most one order of magnitude higher than the rate from cosmic rays at sea level ($10^5$ cm$^{-2}$ per year, roughly corresponding, in a radiation environment generated by a primary hadron source, to a 1 MeV n-eq fluence ten times larger and to a total ionizing dose below 1 mGy yr$^{-1}$). Radiation monitors deployed at various points in the service galleries were used to benchmark the FLUKA calculations [67], obtaining remarkable agreement. In particular, high-energy hadron fluence values measured in the service gallery by two RadMon detectors [68], located in line-of-sight positions at the duct exits at about 50 and 80 m from the target (see Fig. 13), respectively, were reproduced within 10%, with an estimated experimental uncertainty of 20%.

Besides further instances at SPS and LHC energies [69, 70], another example of successful calculation of relevant radiation levels is given by the study of the antiproton decelerator target area [71], where the antiproton beam to be injected into the antiproton decelerator ring is generated by the impact of the Proton Synchrotron proton beam at 26 GeV/c onto an iridium target. Dose values measured on the beam line 10 m downstream of the target (at the station of the PS-ACOL irradiation facility [72]) as well as by a RadMon detector in a quite peripheral position, as indicated in Fig. 14, matched FLUKA results very well. In the first location, 66–80 mGy per pulse of $1.4 \times 10^{13}$ protons compared with a 68–70 mGy prediction. In the RadMon location, a reading of 50 ($\pm$15) Gy over 14 weeks compared with a 60 Gy prediction (see Fig. 14). The same RadMon yields a thermal neutron to high-energy hadron fluence ratio of 5 ($\pm$40%), in full agreement with the simulation outcome of 5 ($\pm$10%).

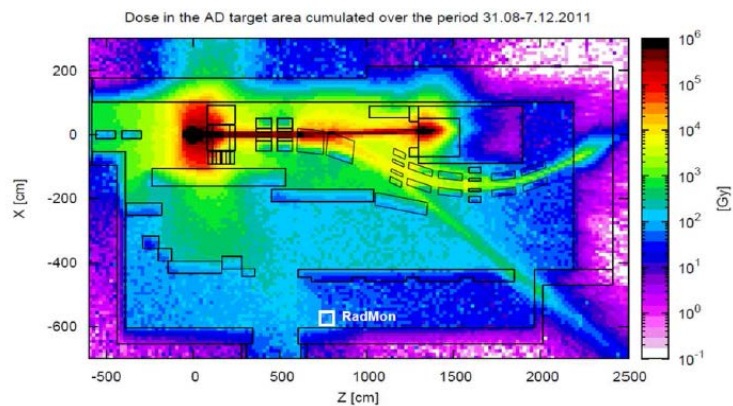

**Fig. 14:** Dose map of the antiproton decelerator target area (top view), as calculated by FLUKA and normalized to the proton beam intensity accumulated over the 14 weeks of the indicated period [71]. Along the missing axis (*Y*), values are averaged over a 10 cm interval, at the height corresponding to the indicated RadMon position. The 26 GeV/*c* proton beam impinges the iridium target at the reference frame origin from the left.

The CNGS incident led to the careful evaluation of all electronic systems located in the LHC underground areas, typically either fully commercial or based on 'commercial-off-the-shelf' components, and of the respective radiation levels. An extensive mitigation strategy, consisting of relocation to safe areas, as well as suitable shielding design and installation, allowed minimization of the single-event-effect impact on the accelerator operation. In fact, single-event-effect-induced downtime decreased from an initial value of 400 h in 2011 to 250 h in 2012, reducing the single-event-effect dump rate referred to cumulated luminosity by a factor of four (from 12 to 3 dumps per inverse femtobarn) [70]. This remarkable achievement has still to be dramatically improved during the new LHC run (Run II) and especially the high-luminosity era. To this purpose, and in the context of any other project implying a challenging radiation environment, a prevention strategy has to be implemented from the early stage onward, entailing the availability of protected areas, possibly relying on a dedicated shielding solution, for electronic equipment not validated by radiation testing, together with the development and adoption of radiation-tolerant and radiation-hardened electronics.

## 3.7 Shielding

In an accelerator context, typical radiation sources are represented by regular and accidental beam impacts on beam-intercepting devices, such as collimators, dumps, targets, or even unexpected obstacles, for instance plastic and metallic dust [73]. In the case of rings, nuclear reactions between beam particles and residual gas nuclei inside the vacuum chamber play an additional role. With electron and positron beams, synchrotron radiation becomes a main concern (it can carry an important power with hadron beams too, but the photon spectrum is much softer and is absorbed by the first material layers). Finally, colliders are affected with beam–beam collision debris around experimental insertions.

For a given source term, the induced radiation levels first depend on the relative position, namely on the radial and longitudinal distance from the shower generation. Clearly, the geometrical attenuation (proportional to the square of the distance in the case of an isotropic source) is often insufficient and, to allow the integration of a radiation facility in the environment, to guarantee accessible areas, or to ensure the correct operation of the electronic equipment (see Section 3.6), a specific shielding has to be designed.

*High energy hadron* propagation is limited (in addition to ionization losses affecting charged species) by non-elastic reactions, replacing the primary particle with lower-energy products. Their occurrence is proportional to the inverse of the nuclear interaction length, i.e., to the density of the traversed material. Therefore, such a radiation field is effectively attenuated by dense materials. When coupled to cost considerations favouring cheap options, this typically suggests the use of iron or bigger volumes of concrete for massive shielding.

At large radial distances from the primary source, beyond a considerable material amount, the radiation field starts to be dominated by *low-energy neutrons*. These are further slowed down by nuclear elastic scattering, more effectively on light nuclei, as in hydrogen-rich materials. Approaching thermal energies, they are removed in the presence of particular isotopes with pronounced capture cross-sections, such as $^{10}$B or $^{113}$Cd. In this context, borated polyethylene is a common solution to wash out the neutron field.

Even in the case of energetic hadron sources, energy deposition is ruled by electromagnetic showers initiated by high-energy *photons* from neutral pion decay. In peripheral areas, photons accompany low-energy neutron propagation, being produced in non-elastic reactions, e.g., radiative capture. With electron beams, bremsstrahlung and synchrotron radiation make photons play a crucial role. At energies above a certain threshold (of the order of 10 MeV in lead and 100 MeV in carbon), photon interaction consists of electron and positron pair production and hence electromagnetic shower development, with further photon generation by lepton bremsstrahlung. Going below that threshold, photons mainly lose energy through Compton scattering, and eventually are absorbed by the photoelectric effect, which is strongly favoured in high-Z materials.

Concerning *high-energy muons*, typically produced in pion and kaon decays, as mentioned at the beginning of this paper, they cannot be stopped within limited distances, since they are affected predominantly by ionization and multiple Coulomb scattering. Bremsstrahlung and direct $e^+e^-$ pair production rule their transport at energies larger than 1 TeV.

For radioprotection purposes, depending on the aspects to be considered, particle fluence in a given location is transformed into *effective dose* or *ambient dose equivalent* (both expressed in sieverts), through the use of respective sets of conversion coefficients, which are a function of particle type and energy [74, 75]. *Prompt* dose equivalent outside a radiation facility, reflecting the relevant radiation level in a public space during normal or accidental operation of the facility, is the quantity to minimize below acceptable limits through the facility shielding design.

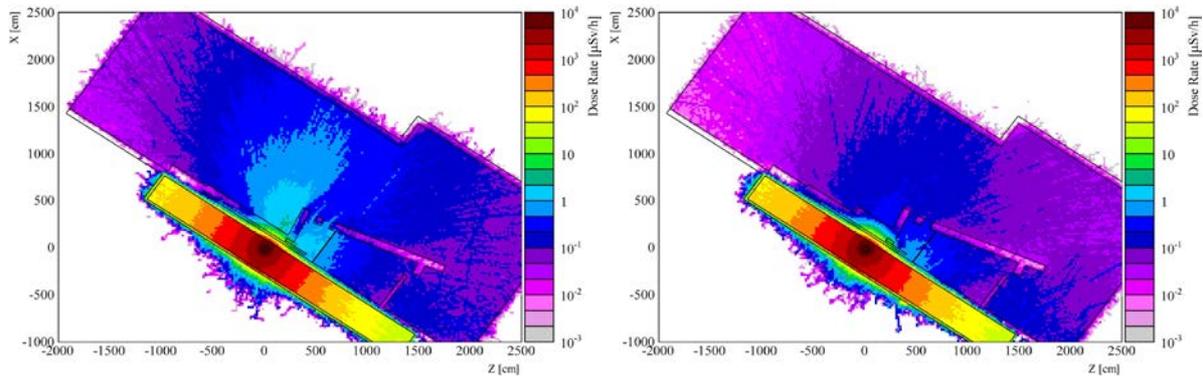

**Fig. 15:** Prompt ambient dose-equivalent maps (top view) in the intersecting storage rings tunnel on the side of the n_TOF EAR-2 neutron beam line, directed along the missing *Y* axis at the origin of the reference frame (courtesy of J. Vollaire). Left: with default shielding, as described in the text. Right: with improved shielding, as described in the text.

As an example, Fig. 15 shows the simulated effect of the shielding optimization performed during the construction of the second experimental area (EAR-2) of the n_TOF facility at CERN [76]. The Proton Synchrotron proton beam impinges at 20 GeV/*c*, with $7 \times 10^{12}$ protons per pulse and an average current of $1.6 \times 10^{12}$ p/s, on the n_TOF spallation target (a massive lead block), generating two neutron beams, which reach EAR-1 and EAR-2, respectively. The first is located after a 185 m horizontal flight path along the proton beam direction, whereas the new area was built at 90°, 20 m above the lead target, to provide a significantly higher neutron flux. Along the path towards EAR-2, the neutron beam line runs contiguous to a tunnel hosting a workplace, just behind a 60 cm thick concrete wall, where strict limits apply in terms of prompt ambient dose equivalent. With default shielding consisting of a concrete block 80 cm thick and 2 m high, neutron streaming induces a dose-equivalent rate exceeding few μSv/h (Fig. 15 left), implying that the area should be classified, from the radioprotection standpoint [77]. Nevertheless, with the extension of the concrete protection, whose volume was increased almost three times, and the addition of two iron plates, a quite significant improvement was achieved, enabling the 0.5 μSv/h limit to be complied with (Fig. 15 right), as nicely confirmed by later radiation monitor measurements (J. Vollaire, private communication).

### 3.8 Activation

As anticipated in Section 2.2, material activation is responsible for continuous delayed emissions, defining the radiation conditions of a facility during its shut-down periods. This also limits access and intervention possibilities during beam absence, and affects equipment handling, e.g., waste disposal. Therefore, activation levels have to be evaluated since the design stage. Calculation reliability mainly depends on accuracy in radionuclide production (see Fig. 5) and on knowledge of actual material compositions. Various activation benchmark experiments have been performed [78–80]. As an example, Fig. 16 (left) shows the activity profile of $^{57}$Co generated along a copper target by a 500 MeV/n $^{238}$U beam [80], penetrating a distance of about 5 mm. As suggested by the measured shape, which is well reproduced by simulation codes, the considered nuclide is mainly produced in secondary neutron re-interactions. Figure 16 (right) illustrates the role of isotopes of different lifetimes. In this case, various samples were put in the vicinity of a copper target irradiated by a 120 GeV/*c* proton and positive pion beam and then transferred to a low-background laboratory, to measure the time evolution of residual dose rates at several distances from the sample [79]. Values scale down with increasing distance, as predicted, and, for the concrete sample considered here, the time profile is shaped by $^{11}$C decay (positron emission with $t_{1/2} = 20$ min) for the very first few hours and by $^{24}$Na decay (β$^-$ transmutation into $^{24}$Mg generating two γ lines with $t_{1/2} = 15$ h) later on.

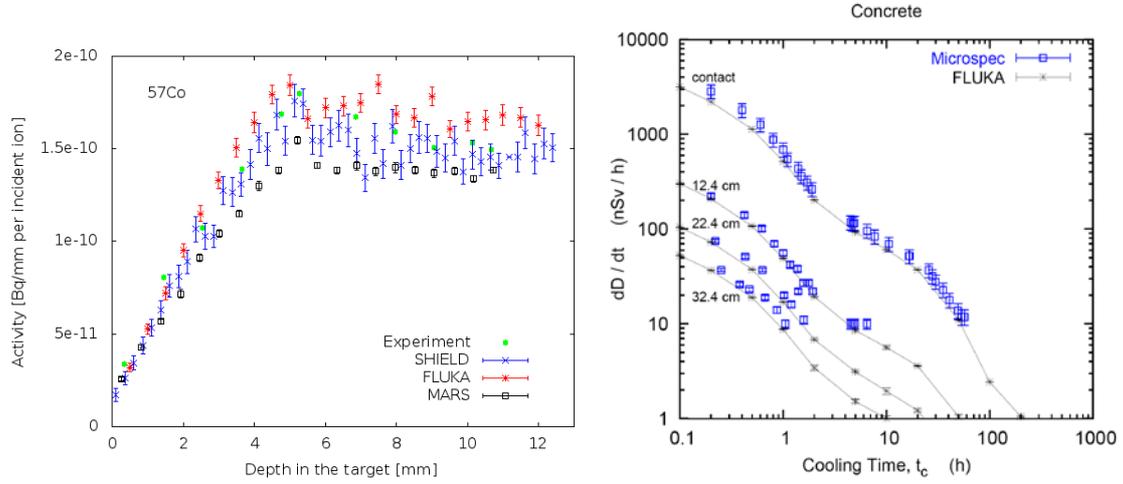

**Fig. 16:** Left: Activity profile of $^{57}$Co in a copper target hit by a 500 MeV/n $^{238}$U beam [80]. Experimental data are compared with the predictions of the indicated codes. Right: Time evolution of residual ambient dose-equivalent rates at different distances from a concrete sample, previously exposed to prompt radiation emerging from a nearby copper target hit by 120 GeV/c protons and positive pions [79]. Blue symbols are experimental data and black symbols connected by lines are FLUKA results.

The calculation of 3D residual dose maps, coupled with an intervention plan detailing the position of the workers and the duration of their actions, allows for the evaluation of individual and collective doses, to be compared with legal limits, design limits (required as facility design criteria not to surpass a given dose per intervention and per year), and optimization thresholds (imposing, if exceeded, optimization of the intervention plan, to minimize the dose to personnel according to the 'as low as reasonably achievable' principle). In this regard, a few guidelines concerning material choice, material amount, and equipment handling, should be considered at the beginning of every new project: whenever possible, lower-activation, radiation-resistant, more easily disposable materials must be preferred; only essential components should be installed, in particular in high-loss areas, and they must be easily accessible and enable fast installation, maintenance, repair, and dismantling.

A special aspect is represented by air activation, which has to be considered from the points of view of release in the environment and accessibility delay of an irradiated area. The activity of a certain air radioisotope inside the latter at the end of the irradiation period $T$ is given by

$$A_T = A_S \left(1 - \exp(-(\lambda + m_{\mathrm{on}})T)\right), \tag{3}$$

where $\lambda$ is the radioisotope's decay probability per unit time and $m_{\mathrm{on}}$ is the relative air exchange rate during irradiation, i.e., the fraction of the total air volume in the area that is renewed per unit time. The quite low interaction probability of particles in air might limit the Monte Carlo statistical accuracy of air radioisotope production; hence, an alternative two-step method involves scoring the energy distribution of hadron fluence in air and then folding it with the cross-sections for radioisotope production from the air target nuclei. In this way, the saturation activity $A_S$ can be calculated as [81]

$$A_S = \frac{V\lambda}{\lambda + m_{\mathrm{on}}} \sum_{P,T,j} \phi_P(E_j) \sigma_{P,T}(E_j) N_T \, (\Delta E)_{j,P}, \tag{4}$$

where the summation has to be performed over the produced hadron species $P$ (mainly neutrons, protons, and charged pions), the air nuclear species $T$ ($^{12}$C, $^{14}$N, $^{16}$O, and $^{40}$Ar), and all the bins $j$ of width $\Delta E$ into which the hadron energy range has been divided. $V$ is the irradiated air volume, $\phi$ is the differential fluence rate, $\sigma$ is the production cross-section for the considered radioisotope, and $N_T$ is the number of target nuclei per unit volume, calculated from the air composition.

For each air radioisotope, the total amount of activity released into the atmosphere during the irradiation period, $T$, is

$$A_{\text{on}} = m_{\text{on}} A_S \left(T - \frac{1-\exp(-(\lambda+m_{\text{on}})T)}{\lambda+m_{\text{on}}}\right) \exp(-\lambda\, t_{\text{on}}) , \qquad (5)$$

where $t_{\text{on}}$ is the time taken by the air flux to reach the release point from the irradiated area where air is being activated. Finally, the total amount of activity released into the atmosphere after shut-down can be obtained by

$$A_{\text{off}} = A_T \frac{m_{\text{off}}}{\lambda+m_{\text{off}}} \exp(-\lambda\, t_{\text{off}}) , \qquad (6)$$

with $m_{\text{off}}$ and $t_{\text{off}}$ representing the same quantities as $m_{\text{on}}$ and $t_{\text{on}}$, respectively, but referred to the period following the irradiation end.

### 3.9 Simulation tools in challenging accelerator applications

The challenge here is to produce a detailed and accurate (to a percent level) model of all particle interactions with 3D system components (up to tens of kilometres of the accelerator lattice in some cases) in the energy region spanning up to 20 decades, as a basis of accelerator, detector, and shielding designs and their performance evaluation, for both short-term and long-term effects.

The current versions of five general-purpose all-particle codes are capable of this: FLUKA, GEANT4, MARS15, MCNP6, and PHITS. These are used extensively worldwide for accelerator applications in conjunction with such accelerator tracking codes as STRUCT [54] and SixTrack [82–84]. A substantial amount of effort (up to several hundreds of person-years) has been put into development of these codes over the last few decades. The user communities for the codes reach several thousands of people worldwide. The five codes listed above can handle a very complex geometry, have powerful user-friendly built-in graphical user interfaces with magnetic field and tally viewers, and variance reduction capabilities. Tallies include volume and surface distributions (1D to 3D) of particle flux, energy, reaction rate, energy deposition, residual nuclide inventory, prompt and residual dose equivalent, number of displacements per target atom for radiation damage, event logs, intermediate source terms, etc. All the aspects of beam interactions with accelerator system components are addressed in sophisticated Monte Carlo simulations benchmarked—wherever possible—with dedicated beam tests.

In accelerator applications, particle shower simulations lie in a multidisciplinary field, in which particle dynamics in accelerators and radiation-matter interaction play together. In fact, their source term is often provided through multiturn tracking in accelerator rings, which requires dedicated codes, eventually dumping, in static loss files, a beam particle sample characterized in the phase space at a certain interface. Conversely, tracking codes happen to be faced with the problem of dealing with particle scattering in beam-intercepting devices, such as collimators. Innovative solutions adopt different types of on-line coupling between tracking and interaction codes, which exchange particle run times to perform their respective tasks. There has been a quantum leap in coupling these general-purpose codes with tracking codes for accelerators:

– MMBLB = MAD-MARS Beam Line Builder (since 2000) [85]. Earlier used the STRUCT code [54] tracking; currently coupled with MAD-X [86, 87], and available in the ROOT-based version [2, 50].

– BDSIM combines C++ in-vacuum accelerator style particle tracking and GEANT4 physics (since the mid-2000s) [88].

– FLUKA LineBuilder and Element Database [89] and active coupling to SixTrack; the two codes communicate with each other through a network port [90].

Figure 17 shows two examples of the complexity affordable nowadays in accelerator line modelling for beam–machine interaction studies, which is coupled to the required accuracy in geometry detail implementation (from vacuum chamber and collimator aperture to magnet coil structure and radiation monitor positioning) as well as in magnetic field treatment and scoring resolution.

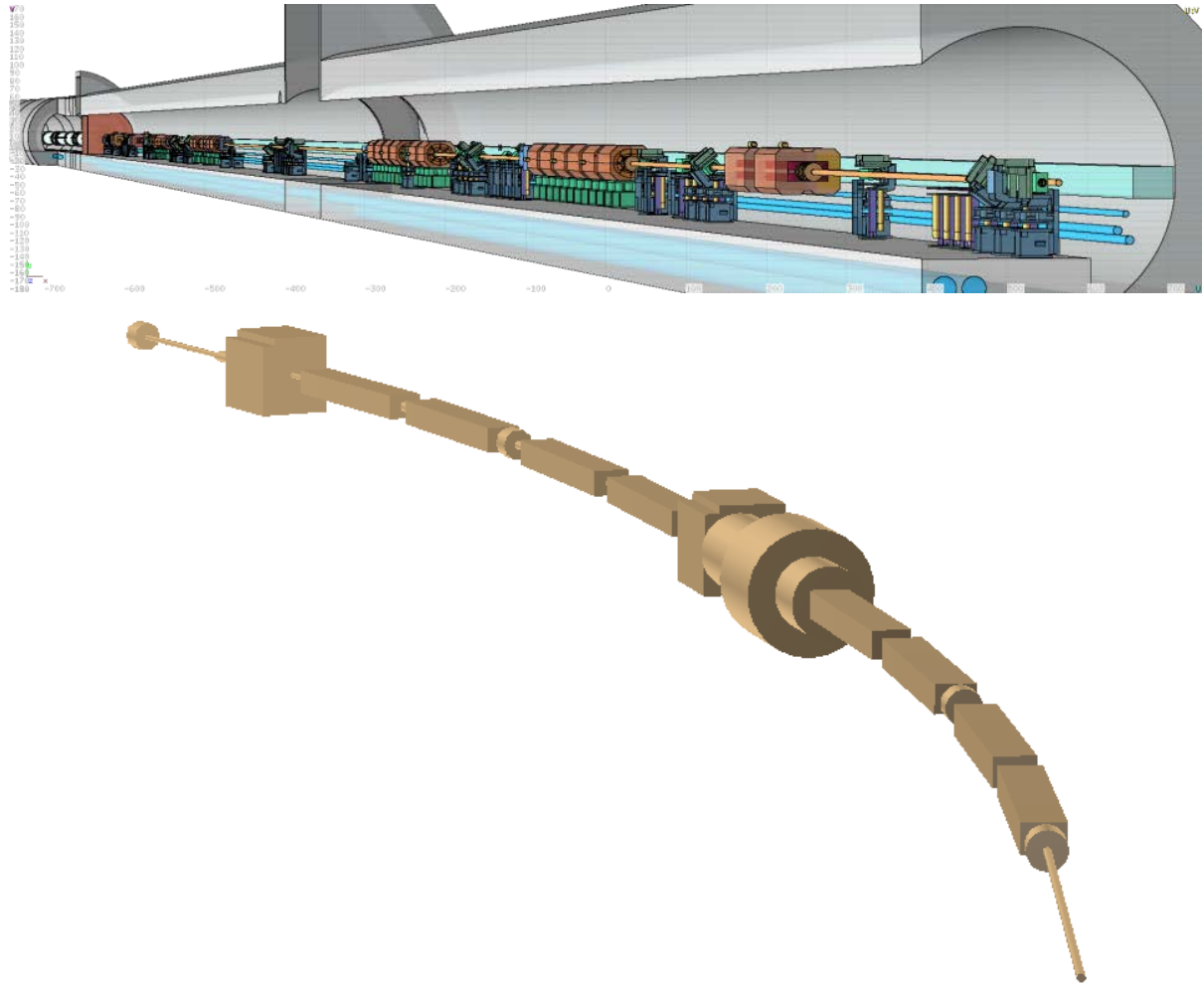

**Fig. 17:** Top: FLUKA geometry of the LHC betatron cleaning insertion (interaction region 7) by the FLUKA LineBuilder and Element DataBase [89]. Bottom: MARS15 geometry of the Fermilab Booster by the ROOT-Based MMBLB [2, 50].

Towards the end of the LHC run 1 (early 2013), several quench tests were performed, to explore the actual quench limits of the superconducting magnets and verify the quality of theoretical calculations [91]. In particular, during a so-called collimation quench test [92, 93] at a beam energy of 4 TeV, the horizontal primary collimator of the betatron cleaning insertion was impacted by a peak proton loss rate equivalent to about 1 MW for 1 s, with no quench occurring in the downstream dispersion suppressor. The propagation of the induced particle shower was measured by the beam loss monitor system [94], giving a picture of the energy deposition profile over several hundred metres, as shown in Fig. 18. This provided a very challenging opportunity to benchmark the adopted SixTrack-FLUKA simulation chain [95], which yielded the impressive agreement reported in the figure, both in terms of pattern and absolute signal comparison, spanning a few orders of magnitude.

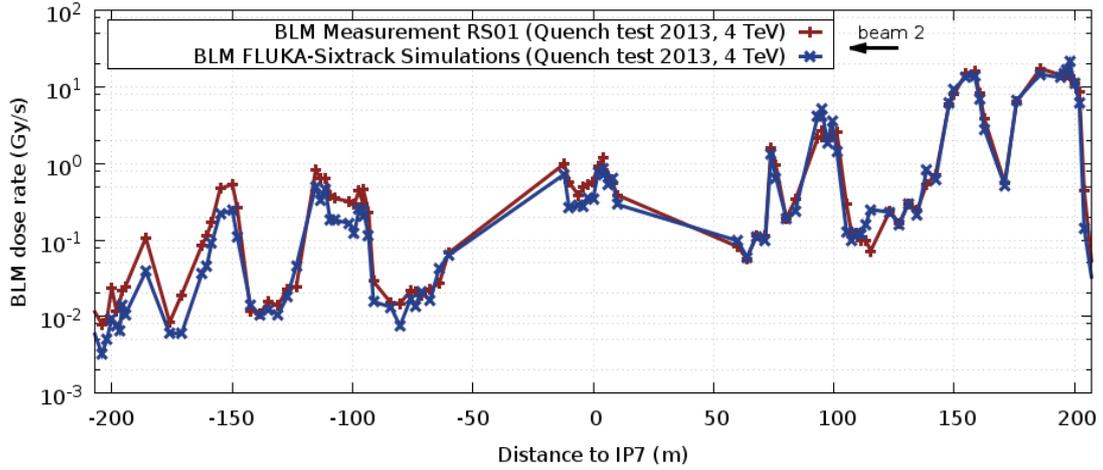

**Fig. 18:** Absolute beam loss monitor signal pattern at the peak loss rate of the 2013 LHC collimation quench test, averaged over the shortest available time interval of 40 µs (RS01): data (red) are compared with predictions (blue) by the SixTrack-FLUKA coupling according to the simulation strategy discussed in Ref. [95].

All the power and the described features of the current version of the MARS15 code are fully exploited in the new neutrino and precision fixed-target experiments for the Intensity Frontier under design and construction in the USA, such as Mu2e and the Long Baseline Neutrino Facility/DUNE. The code was thoroughly benchmarked against the data at corresponding proton beam energies of 8 GeV (for Mu2e) and 120 GeV (for the Long Baseline Neutrino Facility/DUNE). As an example, Fig. 19 shows that MARS15 calculations agree very well with experimental data for a 120 GeV beam on a thick graphite target for pion spectra (left) and neutrino and antineutrino spectra (right) at the MINOS (Main Injector Neutrino Oscillation Search) near detector. Results of the MARS15 calculations for a 2.4 MW beam were used to optimize the design of the Long Baseline Neutrino Facility hadron absorber at the end of the 200 m long decay channel, as well as complicated radiation shielding around the absorber, along with the configuration of all the conventional facilities, as shown in Fig. 20. Note that the calculated power density profiles span 12 orders of magnitude in the region of interest, while for the prompt dose it was necessary to cover 24 decades, which would be impossible without applying the corresponding variance reduction and other sophisticated Monte Carlo techniques.

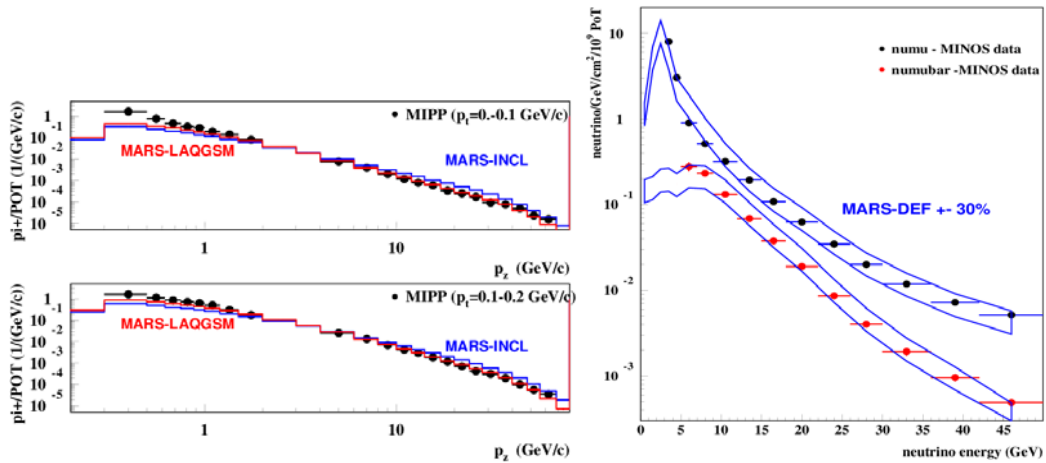

**Fig. 19:** MARS15 results versus recent MIPP (main injector particle production) NuMI target data on (left) pion spectra and (right) MINOS (Main Injector Neutrino Oscillation Search) neutrino and antineutrino spectra at the near detector. Results are normalized per proton on target indicated as POT and PoT.

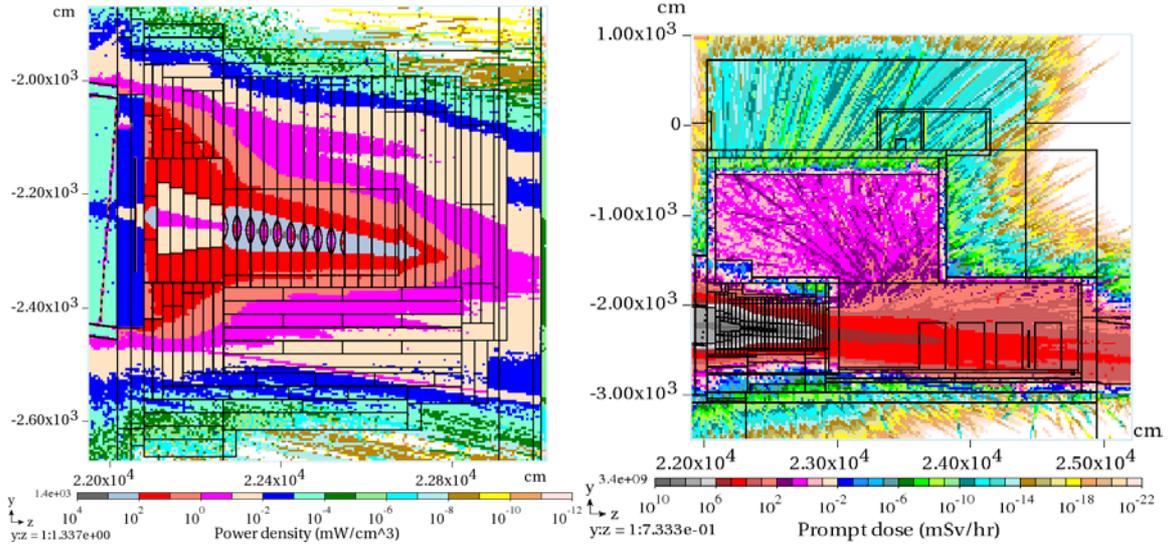

**Fig. 20:** Left: MARS15-calculated power density map in the Long Baseline Neutrino Facility hadron absorber. Right: MARS15-calculated prompt dose distribution in the entire hadron absorber complex.

A very challenging, and at the same time exciting, application was a muon collider project in which the design energy of muon beams varied over several years from 62.5 GeV to 3 TeV. As described in Ref. [96] for the Higgs Factory muon collider, a detailed 3D model was built using MARS15 for the entire collider ring, including the interaction region, the chromaticity correction and matching sections, the arc, the machine-detector interface, and the SiD-like collider detector, with the silicon vertex detector and tracker based on the design proposed for the Compact Muon Solenoid detector upgrade.

Figure 21 shows the 3D model, while Fig. 22 shows the components in the machine-detector interface region. At a muon energy of 62.5 GeV with $2 \times 10^{12}$ muons per bunch, the electrons from muon decays deposit more than 300 kW in the superconducting magnets of the Higgs Factory interaction region and storage ring. This heat deposition corresponds to an unprecedented average dynamic heat load of 1 kW/m around the 300 m long ring, or a multimegawatt room temperature equivalent, if the heat is deposited at helium temperature. That is about one hundred times above acceptable levels. The detector backgrounds in such a project are also much too excessive. The suppression needed on both the fronts has been achieved through substantial effort. First, the lattice and magnets—with a dipole component in the interaction region quadrupole magnets and large apertures varying along the lattice—were designed appropriately. To further protect the collider, thick tungsten masks and liners (with tight elliptical apertures varying according to the beam envelope) in the magnet interconnect regions and inside each magnet have been optimized using massive iterative MARS15 studies. The configuration and composition of a sophisticated tungsten nozzle in the vicinity of the interaction point and other details of the machine-detector interface were optimized simultaneously. As a result, the average dynamic heat load on the superconducting coils of ~1 kW/m was reduced to the allowable value of 10 W/m at 4.5 K, with the peak power density in the coils being reduced to below the quench limit, with a necessary safety margin [96]. The detector backgrounds were also reduced adequately [97].

Various examples of design and operation of ambitious research facilities have been discussed in this paper. All of them required a thorough consideration of the specifics of particle-matter interactions and corresponding physics processes in the phase space regions of interest, understanding of the beam-induced microscopic and macroscopic effects in the components, close iterative work with the lattice, magnet and detector designers, and use of the modern state-of-the-art simulation tools.

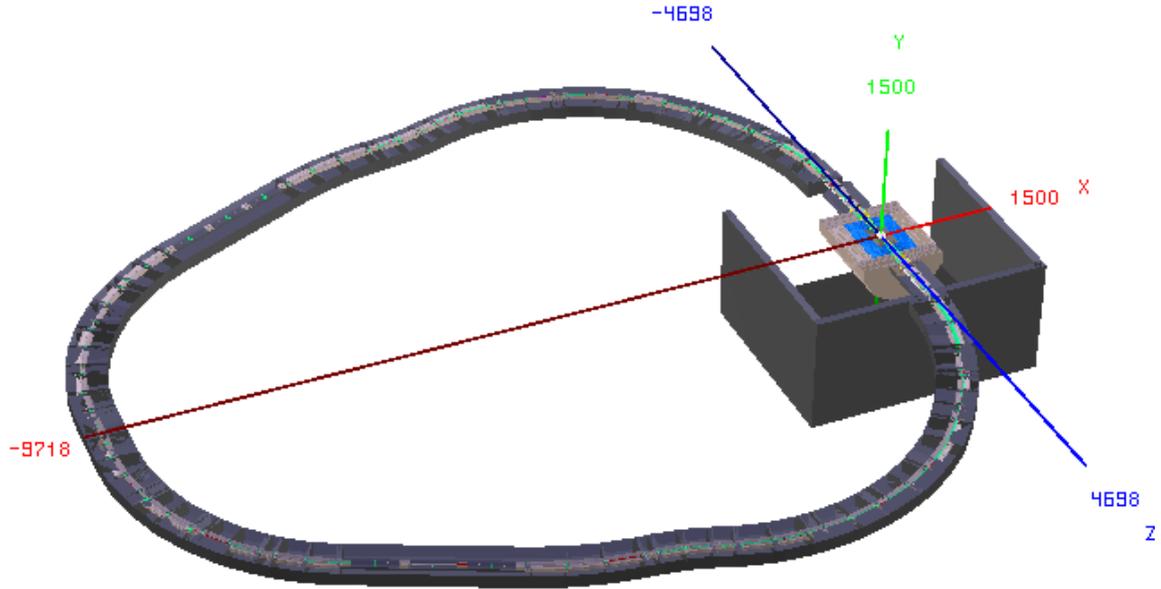

**Fig. 21:** MARS15 3D model of the Higgs Factory muon collider

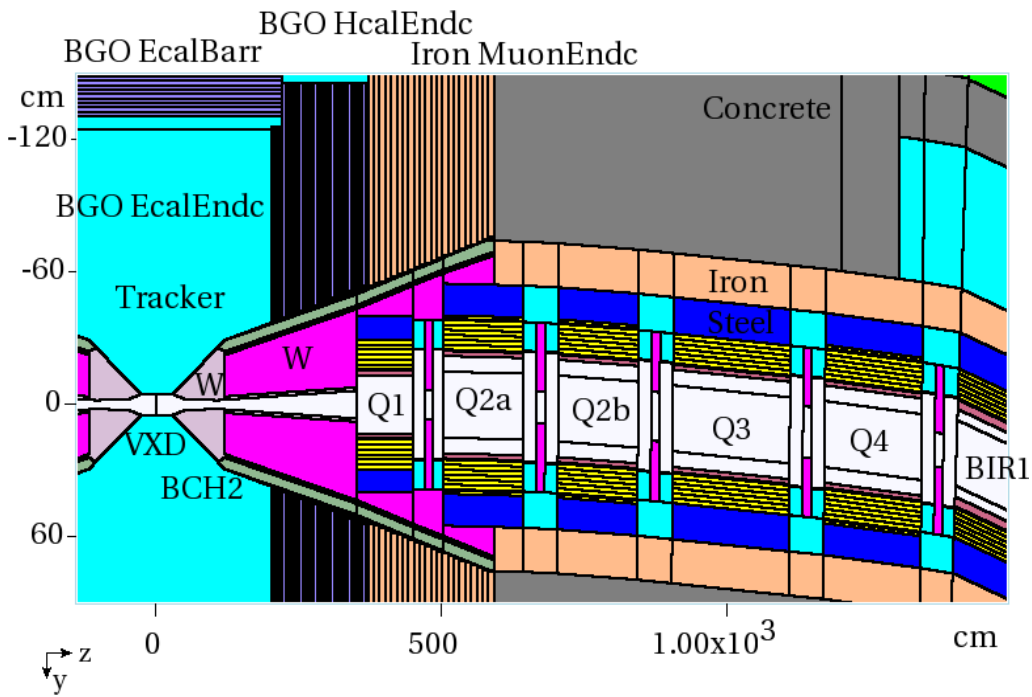

**Fig. 22:** Higgs Factory machine-detector interface MARS15 model with tungsten nozzles on each side of the interaction point, tungsten masks in interconnect regions and tungsten liners inside each magnet [96]. BCH2, borated polyethylene layer; BGO HcalEndc, Bismuth Germanate hadron endcap calorimeter; BGO EcalBarr, Bismuth Germanate electromagnetic barrel calorimeter; BGO EcalEndc, Bismuth Germanate electromagnetic endcap calorimeter; BIR1, first dipole magnet in the interaction region; MuonEndc, muon endcap detector; Q, quadrupole magnets; W, tungsten; VXD, vertex detector.

## Acknowledgements

We wish to thank many CERN and Fermilab colleagues for their very valuable contributions, in particular M. Brugger, A. Ferrari, K. Gudima, I. Rakhno, S. Roesler, S. Striganov, I. Tropin, and V. Vlachoudis.